\documentclass[reprint,amsmath,amssymb,aps,prc,nofootinbib]{revtex4-1}
\usepackage{bm}
\usepackage{graphicx}
\usepackage{dcolumn}
\usepackage{color}
\newcommand{\be}{\begin{equation}}
\newcommand{\ee}{\end{equation}}
\newcommand{\bea}{\begin{eqnarray}}
\newcommand{\eea}{\end{eqnarray}}

\newcommand{\mbss}[1]{_{\mbox{\scriptsize #1}}}

\newcommand{\mbsu}[1]{\mbox{\scriptsize #1}}

\newcommand{\vphu}{\vphantom{*}}
\newcommand{\vphd}{\vphantom{1}}

\newcommand{\ve}{\varepsilon}

\newcommand{\ZRPA}{Z}

\begin{document}

\title{Self-consistency in the phonon space
of the particle-phonon coupling model}

\author{V. Tselyaev}
\email{tselyaev@mail.ru}
\author{N. Lyutorovich}
\affiliation{St. Petersburg State University, St. Petersburg, 199034, Russia}
\author{J. Speth}
\affiliation{Institut f\"ur Kernphysik, Forschungszentrum J\"ulich, D-52425 J\"ulich, Germany}
\author{P.-G. Reinhard}
\affiliation{Institut f\"ur Theoretische Physik II, Universit\"at Erlangen-N\"urnberg,
D-91058 Erlangen, Germany}
\date{\today}

\begin{abstract}
In the paper the non-linear generalization of the time blocking approximation (TBA)
is presented. The TBA is one of the versions of the extended
random-phase approximation (RPA) developed within
the Green-function method and the particle-phonon coupling model.
In the generalized version of the TBA the self-consistency principle is extended onto
the phonon space of the model.
The numerical examples show that this non-linear version of the TBA leads
to the convergence of the results with respect to enlarging the phonon space
of the model.
\end{abstract}


\maketitle

\section{Introduction}
\label{sec:Intr}

The Green-function (GF) method is a powerful tool for solving
the nuclear many-body problem (see Refs. \cite{Migdal67,SWW77}).
General equations of this method are inherently non-linear.
However, in the practical applications, the linearized versions
of these equations are commonly used.
In particular, the linearized Bethe-Salpeter equation for the
response function is equivalent at some additional assumptions
to the well-known random-phase approximation (RPA).
The non-linearity in the nuclear-structure models is frequently
contained only on the mean-field level where
Hartree-Fock (HF) or Hartree-Fock-Bogoliubov (HFB) approximations
and the energy density functional (EDF) theory are applied.
Nevertheless, several approaches were developed where effects of non-linearity
beyond the mean-field level were considered within the so-called
self-consistent RPA
(SCRPA, see Refs. \cite{Schuck1973,Dukelsky1990,Dukelsky1998,Schuck2016})
and within the standard GF method in the Bethe-Salpeter and
in the Dyson equations (see, e.g., Refs.
\cite{Immele1977b,Muraviev1991,Litvinova2015}).

In the mean-field approach, e.g. in the RPA, the model space is restricted to
one-particle--one-hole ($1p1h$) configurations. This is a reliable method to calculate
energies and transition probabilities of low-lying collective states as well as
the mean energies and total strengths of the high-lying giant resonances.
For more details like the spreading width of giant resonances one has to extend
the configuration space by considering e.g. two-particle--two-hole correlations
\cite{Drozdz} or by coupling phonons to the $1p1h$ configurations in a shell model
approach or in a self-consistent way, see Ref.~\cite{Tselyaev2016} and references therein.
In the phonon-coupling models used so far the results depend on the number of phonons
considered and moreover the non-collective states may enter the phonon space that leads
to the violation of the Pauli principle. In a recent publication \cite{Lyutorovich2017}
we presented a scheme to select the most relevant phonons in order to avoid these
shortcomings to some extend. This model is a modification of
the \emph{Time Blocking Approximation} (TBA), Refs.~\cite{TBA89,KTT97,KST04},
which is one of the versions of the extended RPA developed within
the GF method and the particle-phonon coupling model.

In the present paper we develop a non-linear form of the TBA which we name
\emph{Configuration Blocking Approximation} (CBA). This approach is connected
with the self-consistent determination of the phonon space of the model.
It assumes some additional restrictions imposed on the phonons included in this space.
In our study we focus on the analysis of the influence of these restrictions
on the convergence of solutions of the model equations with respect to
enlarging the phonon space.

In Sec.~\ref{sec:RenTBA} the formalism of the non-linear version of the TBA
is introduced. In the first subsection \ref{sec:RPA} we summarize the
conventional self-consistent RPA which is the basis for all extended
models. Our new approach in its general form
is presented in subsection \ref{sec:nTBA}. On
a first glance the equations look identical to the TBA
equations. There is indeed only one decisive difference: The phonons
which couple to the single-particle propagators are not the solutions
of the RPA equation but the solutions of the TBA itself which
introduces the non-linearity. The first modification concerns the CBA
which is presented in Sec.~\ref{sec:CBA}.
Here we augment the non-linear model
with additional restrictions imposed on the phonon space.
The reduced form of this non-linear model is presented in
Sections \ref{sec:CBAdiag} and \ref{sec:CBAjust}.  The method of
construction of the phonon space of the model is described in
Sec.~\ref{sec:RPAspace}. In the short subsection \ref{sec:TBAlim} the
relation between CBA and TBA is explicitly demonstrated.  In
Sec.~\ref{sec:res} numerical results of our new approach are
presented. We compare previous results with the present ones with
the emphasis on the dependence on the size of the phonon configuration
space. The conclusions are given in the last Section.

\section{The formalism of non-linear TBA }
\label{sec:RenTBA}

\subsection{RPA as starting point and basis}
\label{sec:RPA}

Before presenting the involved TBA theories, we summarize here the
RPA which serves as starting point for the further development and
provides an appropriate basis of the description. Doing so, we
introduce ``en passant'' also the basic notations used henceforth.

RPA determines the excitation spectrum of a many-body system in the
$1p1h$ vicinity of the ground state. There are several ways to write
the RPA equations. We use here a formulation in terms of the
response operator $R^{\mbss{RPA}}_{\vphd}\equiv{R}^{\mbss{RPA}}_{12,34}$
which plays the role of a one-body operator in $1p1h$ space
(the numerical indices stand for the
sets of the quantum numbers of some single-particle basis).
This then reads
\begin{subequations}
\begin{eqnarray}
  R^{\mbss{RPA}}_{\vphd}(\omega)
  &=&
  -\bigl(\,\omega - \Omega^{\mbss{RPA}}_{\vphd}\bigr)^{-1}
  M^{\mbss{RPA}}_{\vphd},
\label{rfrpa}
\\
  \Omega^{\mbss{RPA}}_{12,34}
  &=&
  \Omega^{(0)}_{12,34}
  + \sum_{56} M^{\mbss{RPA}}_{12,56}\,{V}^{\vphu}_{56,34}\,,
\label{omrpa}
\\
  \Omega^{(0)}_{12,34}
  &=&
  h^{\vphu}_{13}\,\delta^{\vphu}_{42} -
  \delta^{\vphu}_{13}\,h^{\vphu}_{42}\,,
\label{omrpaz}
\\
  M^{\mbss{RPA}}_{12,34}
  &=&
  \delta^{\vphu}_{13}\,\rho^{\vphu}_{42} -
  \rho^{\vphu}_{13}\,\delta^{\vphu}_{42}\,,
\label{mrpa}
\\
  h^{\vphu}_{12}
  &=&
  \frac{\delta E[\rho]}{\delta\rho^{\vphu}_{21}}\,,
\\
  {V}^{\vphu}_{12,34}
  &=&
 \frac{\delta^2 E[\rho]}
 {\delta\rho^{\vphu}_{21}\,\delta\rho^{\vphu}_{34}}\,,
\label{screl}
\end{eqnarray}
\end{subequations}
where $E[\rho]$ is the EDF of the model, the $\Omega$ are Hamiltonian
matrices in $1p1h$ space (thus carrying four indices),
$M^{\mbss{RPA}}_{\vphd}$ is the RPA norm matrix,
$\rho$ is the single-particle density matrix,
$h$ is the single-particle Hamiltonian,
and $V$ is the residual interaction in the particle-hole channel.
In the following we will also use the indices $p$ and $h$ to
label the single-particle states of particles and holes,
respectively, in the basis in which the density matrix $\rho$ and
the Hamiltonian $h$ are diagonal
(so that $h^{\vphu}_{pp}=\ve^{\vphu}_{p}$ and $h^{\vphu}_{hh}=\ve^{\vphu}_{h}$).
The poles of $R^{\mbss{RPA}}_{\vphd}(\omega)$ determine the RPA spectrum of
eigenfrequencies $\omega^{\vphu}_{n}$.

The RPA response operator can be expressed explicitly in terms of the
spectral representation
\begin{equation}
  R^{\mbss{RPA}}_{\vphd}(\omega)
  =
  -\sum_{n} \frac{\,\sigma^{\vphu}_{n}|\,\ZRPA^{n}\rangle
                    \langle \ZRPA^{n}|}
                 {\omega - \omega^{\vphu}_{n}}\;,
\label{rfrpaexp}
\end{equation}
where
$\sigma^{\vphu}_{n}=\mbox{sgn}(\omega^{\vphu}_{n})$,
the $|\,\ZRPA^{n} \rangle$ are the $n$-th RPA eigenstate
whose details are given by the explicit $1p1h$ expansion coefficients
$\ZRPA^{n}_{12}$.
Inserting the spectral expansion into the response equation
(\ref{rfrpa}) yields the RPA equations explicitly in terms
of the expansion coefficients
\begin{equation}
  \sum_{34} \Omega^{\mbss{RPA}}_{12,34}\,\ZRPA^{n}_{34}
  =
  \omega^{\vphu}_{n}\,\ZRPA^{n}_{12}\,.
\label{rpaeve}
\end{equation}
Both forms, (\ref{rfrpa}) as well as (\ref{rpaeve}), are used in
practice for the practical solution of the RPA equations.

The entity of all RPA eigenstates constitutes an expansion in $1p1h$
space
which is ortho-normal
\begin{subequations}
\begin{equation}
  \langle\,\ZRPA^{n'}\,|\,M^{\mbss{RPA}}_{\vphd}|\,\ZRPA^{n}\rangle
  =
  \delta_{nn'}^{\vphu}  \mbox{sgn}(\omega^{\vphu}_{n})
\label{zmz}
\end{equation}
and complete according to the closure relation
\begin{equation}
  \sum_{n}\sigma^{\vphu}_{n}|\,\ZRPA^{n} \rangle \langle \ZRPA^{n}|
  =
  M^{\mbss{RPA}}_{\vphd}
  \;.
\label{rpacr}
\end{equation}
\end{subequations}
The RPA basis does also allow a complete representation of matrix
operators. Any matrix $A\equiv A^{\vphu}_{12,34}$ in the $1p1h$ space can
be written as
\begin{subequations}
\label{eq:reprA}
\begin{eqnarray}
  A
  &=&
  \sum_{nn'}|\,\ZRPA^{n} \rangle A^{\vphu}_{nn'} \langle \ZRPA^{n'}|\;,
\label{rpaexp}
\\
  A^{\vphu}_{nn'}
  &=&
  \sigma^{\vphu}_{n}\sigma^{\vphu}_{n'}\,
  \langle \ZRPA^{n}|M^{\mbss{RPA}}_{\vphd} A\,M^{\mbss{RPA}}_{\vphd}
  |\,\ZRPA^{n'}\rangle\;.
\label{rnndef}
\end{eqnarray}
\end{subequations}

\subsection{The non-linear TBA equations}
\label{sec:nTBA}

The RPA equations were forcefully closed by the quasi-boson
approximation \cite{Rowebook,RingSchuck}. Releasing this restriction,
they will couple to higher configurations which can be efficiently
expanded in terms of $1p1h\otimes$phonon configuration, where
``phonon'' stands for a subset of RPA eigenstates which have large
collective strength and so couple most strongly to the pure $1p1h$
states in the expansion. An explicit expansion in the full
$1p1h\otimes$phonon space is extremely costly because it includes too
many unimportant contributions.  The phonon-coupling models simplify
that task by maintaining a $1p1h$ expansion while including the
temporary detours through $1p1h\otimes$phonon space by modifying the
RPA interaction matrix. We use the phonon-coupling model here in the
form of the TBA \cite{TBA89,KTT97,KST04}. In fact, we consider it
here, in extension of previous applications, in its self-consistent
form. This gives rise to a non-linear equation for the eigenstates $\nu$ in terms
of $1p1h$ coefficients $z^{\nu}_{12}$ and eigenfrequency
$\omega^{\vphu}_{\nu}$ which reads
\begin{subequations}
\label{eq:SCTBA}
\begin{equation}
  \sum_{34} \Omega^{\mbss{CBA}}_{12,34}
  (\omega^{\vphu}_{\nu})\,z^{\nu}_{34}
  =
  \omega^{\vphu}_{\nu}\,z^{\nu}_{12}
  \,,
  \label{tbaeve}
\end{equation}
with
\begin{eqnarray}
  \Omega^{\mbss{CBA}}_{12,34}(\omega)
  &=&
  \Omega^{\mbss{RPA}}_{12,34}
  +\sum_{56} M^{\mbss{RPA}}_{12,56}\,\bar{W}^{\vphu}_{56,34}(\omega)
  \,,
\label{omtba}
\\
  \bar{W}^{\vphu}_{12,34}(\omega)
  &=&
  {W}^{\vphu}_{12,34}(\omega) - {W}^{\vphu}_{12,34}(0)
  \,.
\label{wsub}
\end{eqnarray}
\end{subequations}
The new ingredient in the non-linear TBA
(the superscript CBA in the matrix $\Omega^{\mbss{CBA}}_{\vphd}(\omega)$
will be explained in Sec.~\ref{sec:CBA})
is the matrix $\bar{W}(\omega)$ which
represents the induced interaction generated by the intermediate
$1p1h\otimes$phonon configurations.  The subtraction of $W(0)$ in
Eq.~(\ref{wsub}) is necessary to avoid perturbation of the mean-field
ground state \cite{Toe88,Gue93} and to ensure stability of solutions
of the TBA eigenvalue equation (see \cite{Tselyaev2013}).  The key
point is that ${W}(\omega)$ is expanded in terms of the
$1p1h\otimes$phonon configurations as
\begin{subequations}
\label{eq:Winduc}
\begin{eqnarray}
  {W}^{\vphu}_{12,34}(\omega)
  &=&
  \sum_{c,\;\sigma}\,\frac{\sigma\,{F}^{c(\sigma)}_{12}
  {F}^{c(\sigma)*}_{34}}
  {\omega - \sigma\,\Omega^{\vphu}_{c}}
  \,,
\label{wdef}
\\
  \Omega^{\vphu}_{c}
  &=&
  \ve^{\vphu}_{p'} - \ve^{\vphu}_{h'} + \omega^{\vphu}_{\nu}
  \,,
  \quad \omega^{\vphu}_{\nu}>0
\,,
\label{omcdef}
\end{eqnarray}
\end{subequations}
where $\sigma = \pm 1$, $\,c = \{p',h',\nu\}$ is an combined index for the
$1p1h\otimes$phonon configurations, $\nu$ is the phonon's index.
We emphasize that these are the TBA phonons as they emerge from
the fully fledged TBA equation.
These phonons enter the induced interaction through the $F$ amplitudes
\begin{subequations}

\begin{eqnarray}
  {F}^{c(+)}_{ph}
  &=&
  \delta^{\vphu}_{pp'}\,g^{\nu}_{h'h}\!-\!
  \delta^{\vphu}_{h'h}\,g^{\nu}_{pp'}
  \,,
\label{fcdef}
\\
  g^{\nu}_{12}
  &=&
  \sum_{34} {V}^{\vphu}_{12,34}\,{z}^{\nu}_{34}
  \,,
\label{gndef}
\end{eqnarray}
which obey the symmetry relations
\begin{equation}
  {F}^{c(-)}_{12}={F}^{c(+)*}_{21},\qquad
  {F}^{c(-)}_{ph}={F}^{c(+)}_{hp}=0
  \,.
\label{fcrel}
\end{equation}
\end{subequations}

It is important to note that the larger expansion space which is
implicitly contained in TBA has an impact on the normalization of the
$z$ coefficients.  The RPA norm (\ref{zmz}) is extended to
\begin{subequations}
\begin{eqnarray}
  \langle\,{z}^{\nu}\,|\,M^{\mbss{RPA}}_{\vphd} -
  W^{\prime}_{\nu}\,|\,{z}^{\nu} \rangle
  &=&
  \mbox{sgn}(\omega^{\vphu}_{\nu})
  \,,
\label{zmwz}
\\
  W^{\prime}_{\nu}
  &=&
  \biggl(\frac{d\,W(\omega)}{d\,\omega}\biggr)_{\omega\,=\,\omega_{\nu}}
  \,.
\label{wnn2}
\end{eqnarray}
\end{subequations}
 The second term $\propto W^{\prime}_{\nu}$ is not connected with the non-linearity
 of the TBA but arises already in the conventional TBA.
 It accounts for the probability carried over to the space of $1p1h\otimes$phonon
configurations. Correspondingly, the first term, covering the content
of pure $1p1h$ states, is relatively reduced.
These non-linear TBA equations are self-consistent because the
${z}^{\nu}_{\vphd}$ amplitudes
and the energies $\omega^{\vphu}_{\nu}$
which emerge from the eigenvalue
equation (\ref{tbaeve}) are fed back into the phonon coupling
amplitudes ${F}$ and the energies $\Omega^{\vphu}_{c}$.
This approach is thus superior to standard TBA and
can be considered as the first iteration toward the full scheme.
However, self-consistency involves subtle complications which inhibit
immediate, naive solution of the
Eqs. (\ref{eq:SCTBA},\ref{eq:Winduc}). The problem is solved by
configuration blocking outlined in the next subsection.  This delivers
at the same time as extra bonus an unambiguous and very efficient rule
for confining the intermediate states to the most relevant phonons.

\subsection{Configuration blocking approximation (CBA)}
\label{sec:CBA}

In the non-linear TBA described above the following contradiction
arises. Equations of the ordinary TBA include configurations of the
type $1p1h$ and $1p1h\otimes$phonon, where the phonons are determined
within the RPA.  If we replace the RPA phonons in the matrix
$W(\omega)$ by the solutions of the TBA equation (\ref{tbaeve}), the
resulting equations will include implicitly configurations of the
type $3p3h$ and higher (see \cite{TBA89,KTT97,KST04}). This will be
reflected in the spectrum of the TBA solutions which acquires the huge
spectral density of complex configurations.  Feeding this back into
the induced interaction $W$ grows intractable. And, more important, it
becomes contradictory as configurations of this type goes beyond the framework of the TBA.
Already at the level of the $1p1h$ configurations of RPA, we have to select the few most collective
phonons to render TBA manageable and consistent (see section
\ref{sec:RPAspace}). But now we obtain a swarm of states which have
more strength in higher configurations, i.e. in the $W^\prime$ term in
the norm (\ref{zmwz}), than in its $1p1h$ head. These states are
clearly to  be excluded.
To formalize this decision,
let us rewrite the normalization (\ref{zmwz}) in the form
\begin{subequations}
\begin{eqnarray}
  &&(z^{\nu})^2_{\mbsu{RPA}} + (z^{\nu})^2_{\mbsu{CC}}
  =
  1
\,,
\label{zmwz2}
\\
  &&(z^{\nu})^2_{\mbsu{RPA}}
  =
  \mbox{sgn}(\omega^{\vphu}_{\nu})\,
  \langle\,{z}^{\nu}\,|\,M^{\mbss{RPA}}_{\vphd}|\,{z}^{\nu} \rangle\,,
\label{zzrpa}
\\
  &&(z^{\nu})^2_{\mbsu{CC}}
  =
  -\mbox{sgn}(\omega^{\vphu}_{\nu})\,
  \langle\,{z}^{\nu}\,|\,W^{\prime}_{\nu}\,|\,{z}^{\nu} \rangle
\,.
\label{zzcc}
\end{eqnarray}
\end{subequations}
The term $(z^{\nu})^2_{\mbsu{RPA}}$ in Eq. (\ref{zmwz2}) represents
the contribution of the $1p1h$ (RPA) components to the norm.  The term
$(z^{\nu})^2_{\mbsu{CC}}$ represents the contribution of the complex
configurations.  It is obvious that all states with dominant
$(z^{\nu})^2_{\mbsu{CC}}$ must be discarded from the set of TBA
phonons because they cannot contain any more sufficient collectivity
to contribute to the induced interaction.  To block these
contributions which are dominated by complex configurations from
entering the phonon space of TBA we impose the condition
\begin{subequations}
\begin{equation}
(z^{\nu})^2_{\mbsu{RPA}} > (z^{\nu})^2_{\mbsu{CC}}\,,
\label{czz1}
\end{equation}
which together with Eq. (\ref{zmwz2}) means that
\begin{equation}
(z^{\nu})^2_{\mbsu{RPA}} > \zeta^2_{\mbsu{min}}
\;,\quad
 \zeta^2_{\mbsu{min}}=\frac{1}{2}
\;.
\label{czz2}
\end{equation}
\end{subequations}
We introduce the parameter $\zeta^2_{\mbsu{min}}$ to make the impact
of the condition (\ref{czz2}) visible throughout the formalism.  Only
TBA states which satisfy Eq.~(\ref{czz2}) will be included into the
induced interaction of TBA. We refer to this model as the
configuration blocking approximation (CBA). It is a combination of
non-linear TBA and norm blocking determined
by Eq.~(\ref{czz2}).

The blocking condition (\ref{czz2}) can be formulated in terms of a
blocking factor $f_{\nu}$ which we have to introduce into Eq.~(\ref{fcdef})
which now reads as:
\begin{subequations}
\label{eq:fcph}
\be
  {F}^{c(+)}_{ph} =
  f_{\nu}\left(\delta^{\vphu}_{pp'}\,g^{\nu}_{h'h}\!-\!
  \delta^{\vphu}_{h'h}\,g^{\nu}_{pp'}\right)
  \,.
\label{fcdefa}
\ee
To automatically embody the blocking condition we
put
\begin{equation}
  f^2_{\nu}
  =
  \theta \bigl((z^{\nu})^2_{\mbsu{RPA}} - \zeta^2_{\mbsu{min}}\bigr)
  \,,
\label{fcutdef}
\end{equation}
\end{subequations}
where $\theta$ is the Heaviside step function.
Thus, in a sense, one can consider $f^2_{\nu}$ as an occupation number
for the phonons.

At this point, however, the following difficulty arises.  The TBA
equations (\ref{eq:SCTBA}--\ref{eq:Winduc}) combined with blocking
condition (\ref{czz2}) pose a highly non-linear problem. It is thus
not guaranteed that a unique solution exists.  In fact, one will find
a couple of solutions. In the spirit of dominance of $1p1h$ states, we
select as the most wanted solution the one which maximizes the total
$1p1h$ content of the TBA {\it active} phonons
[i.e. the phonons which enter the matrix $W(\omega)$].
We recall that $1p1h$ content
is defined as the contribution $(z^{\nu})^2_{\mbsu{RPA}}$ to the norm
(\ref{zmwz}). Thus we impose additionally the criterion that
we select that TBA solution which yields
\begin{equation}
\sum_{\nu_a} (z^{\nu_a})^2_{\mbsu{RPA}} = \mbox{max}
\label{sumza}
\end{equation}
where the summation runs over the TBA active phonons only,
i.e. those states $\nu_a$ which obey condition  (\ref{czz2}).

\subsection{CBA in diagonal approximation}
\label{sec:CBAdiag}

Even with the above discussed restrictions imposed on the space of the
TBA phonons, the exact solution of the system of Eqs.
(\ref{eq:SCTBA}), (\ref{eq:Winduc}), (\ref{eq:fcph}),
(\ref{gndef}), (\ref{fcrel}), and (\ref{zzrpa})
remains a rather difficult task.
To simplify these equations further we make use of a diagonal
approximation to the induced interaction
$\bar{W}(\omega)$. Similar as in case of RPA,
also the CBA response function
\begin{subequations}
\begin{equation}
  R^{\mbss{CBA}}_{\vphd}(\omega)
  =
  -\bigl[\,\omega - \Omega^{\mbss{CBA}}_{\vphd}(\omega)\,\bigr]^{-1}
  M^{\mbss{RPA}}_{\vphd}
\label{rftba}
\end{equation}
has a spectral representation
\begin{equation}
  R^{\mbss{CBA}}_{\vphd}(\omega)
  =
  -\sum_{\nu} \frac{\,\sigma^{\vphu}_{\nu}
  |\,z^{\nu} \rangle \langle z^{\nu}|}
  {\omega - \omega^{\vphu}_{\nu}}
  \;,
\label{rfcbaexp}
\end{equation}
\end{subequations}
where $\sigma^{\vphu}_{\nu}=\mbox{sgn}(\omega^{\vphu}_{\nu})$.
As stated in section \ref{sec:RPA},
any matrix $A$ in the $1p1h$ space can be written in terms of the
RPA amplitudes $\ZRPA^{n}_{12}$ according to
representation (\ref{eq:reprA}). We apply that to the CBA response
matrix.
First, we recast the definition (\ref{rftba}) of the response
operator to a defining equations
\begin{equation}
  \bigl[\,\Omega^{\mbss{CBA}}_{\vphd}(\omega) - \omega\,\bigr]
  R^{\mbss{CBA}}_{\vphd}(\omega)
  =
  M^{\mbss{RPA}}_{\vphd}
\label{rftbaeq}
\end{equation}
and write it explicitly as matrix equation in the basis
of RPA states
\begin{eqnarray}
  &&\sum_{n''}\bigl[(\omega - \omega^{\vphu}_{n})\,\delta^{\vphu}_{n,n''}
- \sigma^{\vphu}_{n}\bar{W}^{\vphu}_{nn''}(\omega)\bigr]\,
  R^{\mbss{CBA}}_{n''n'}(\omega)
\nonumber\\
  &&= -\sigma^{\vphu}_{n}\,\delta^{\vphu}_{n,n'}\,,
\label{rfnneq}
\\
 &&\bar{W}^{\vphu}_{nn'}(\omega) =
\langle
\ZRPA^{n}|\,\bar{W}^{\vphu}_{\vphd}(\omega)\,|\ZRPA^{n'}
\rangle
\,.
\label{wnndef}
\end{eqnarray}
This equation now is solved in diagonal approximation
yielding the approximate response
\begin{equation}
  {\tilde R}^{\mbss{CBA}}_{nn'}(\omega)
  =
  - \frac{\sigma^{\vphu}_{n}\,\delta^{\vphu}_{n,n'}}
  {\omega - \omega^{\vphu}_{n}
  - \sigma^{\vphu}_{n}\bar{W}^{\vphu}_{nn}(\omega)}
  \,.
\label{sold1}
\end{equation}

Furthermore, we note that the matrix element
$\bar{W}^{\vphu}_{nn}(\omega)$ is composed, according to
Eqs. (\ref{eq:Winduc}), as a sum of two terms
\begin{equation}
  \bar{W}^{\vphu}_{nn}(\omega)
  =
  \sum_{\sigma = \pm 1}
  \bar{W}^{(\sigma)}_{nn}(\omega)
  \;.
\label{wnnsum}
\end{equation}
It is known already from RPA that the terms with negative frequency
are very small for stable ground states \cite{Rowebook} and only those
situations are considered here.
We assume that these negative-frequency contributions are
not larger than off-diagonal terms which are omitted in the
diagonal approximations and neglect them altogether.
According to Eq.~(\ref{fcrel}), it means that the term
$\bar{W}^{(\sigma)}_{nn}(\omega)$ with $\sigma = - \sigma^{\vphu}_{n}$
can be neglected.
Thus we assume
\begin{equation}
  \bar{W}^{\vphu}_{nn}(\omega)
  =
  \bar{W}^{(\sigma^{\vphu}_{n})}_{nn}(\omega)
  \,.
\label{wndef}
\end{equation}

\subsection{Justification of the blocking value $\zeta^2_{\mbsu{min}}$}
\label{sec:CBAjust}

The diagonal approximation simplifies the mathematical structure of
the response poles to an extend that we can substantiate the choice
$\zeta^2_{\mbsu{min}}=1/2$ as done in condition (\ref{czz2}) on formal
grounds. For the following derivations, we employ that the TBA
corrections are small as compared to the leading RPA structure and
label specifically $\nu\rightarrow(n,q)$ where $n$ stands for a
certain RPA state which becomes ``bandhead'' of the subsequent TBA
structure and $q$ labels the many sub-states in the structure.
Skipping the sum over $\sigma$ in Eq. (\ref{wnnsum}) simplifies the
structure of the response function
${\tilde{R}}^{\mbss{CBA}}_{nn}(\omega)$ such that its poles
$\tilde{\omega}^{\vphu}_{n,q}$ become the roots of the equation
\begin{subequations}
\label{poleq3}
\begin{equation}
  \sigma^{\vphu}_n\tilde{\omega}^{\vphu}_{n,q}\biggl[ 1 + \sum_{c}
  \frac{|{\tilde F}^{c(\sigma_n)}_{n}|^2}
  {{\tilde \Omega}^{\vphu}_{c}\,
  ({\tilde \Omega}^{\vphu}_{c}
  -
  \sigma^{\vphu}_n\tilde{\omega}^{\vphu}_{n,q})}\biggr]
  =
  |\,\omega^{\vphu}_{n}|\,,
\label{poleq}
\end{equation}
\bea
{\tilde F}^{c(\sigma)}_{n} &=& \sum_{12} Z^{n*}_{12} F^{c(\sigma)}_{12},
\label{fcndef}\\
  {\tilde \Omega}^{\vphu}_{c}
  &=&
  \ve^{\vphu}_{p'} - \ve^{\vphu}_{h'} +
  \sigma^{\vphu}_{n'} \tilde{\omega}^{\vphu}_{n',q'}
\label{tomega}
\eea
\end{subequations}
where we use the combined index $c\!\equiv\!(p',h',\nu'\!=\!(n',q'))$.
From that it follows that
$\sigma^{\vphu}_{n}\tilde{\omega}^{\vphu}_{n,q} > 0$ for all $n$ and
$q$.  Now, the pole expansion of the (diagonal) response (\ref{sold1})
has the form
\begin{subequations}
\begin{eqnarray}
  {\tilde R}^{\mbss{CBA}}_{nn'}(\omega)
  &=&
  - \sum_{q}
  \frac{\sigma^{\vphu}_{n}\,\zeta^2_{n,q}
  \,\delta^{\vphu}_{n,n'}}
  {\omega - \tilde{\omega}^{\vphu}_{n,q}}
  \,,
\label{rdexp}
\\
  \zeta^2_{n,q}
  &=&
  \biggl[ 1 + \sum_{c}
  \frac{|{\tilde F}^{c(\sigma_n)}_{n}|^2}
  {({\tilde \Omega}^{\vphu}_{c} - \sigma^{\vphu}_n\tilde{\omega}^{\vphu}_{n,q})^2}
  \biggr]^{-1}
  \,.
\label{andef}
\end{eqnarray}
\end{subequations}
This allows to derive a sum rule for the coefficients $\zeta^2_{n,q}$
 by comparing the first terms of the expansions
in powers of $1/\omega$ of the right-hand sides of Eqs. (\ref{sold1})
and (\ref{rdexp}). Namely, we have for all $n$:
\begin{equation}
  \sum_{q}\zeta^2_{n,q} = 1\,.
\label{ansum}
\end{equation}
From Eqs. (\ref{rpaexp}) and (\ref{rdexp}) we obtain
\begin{equation}
{\tilde R}^{\mbsu{CBA}}(\omega) = - \sum_{n,q}
\frac{\sigma^{\vphu}_{n}\,\zeta^2_{n,q}
\,|\,\ZRPA^{n} \rangle \langle \ZRPA^{n}|}
{\omega - \tilde{\omega}^{\vphu}_{n,q}}\,.
\label{rfrentbaexp}
\end{equation}
Comparing Eqs. (\ref{rfcbaexp}) and (\ref{rfrentbaexp}), we confirm
that in the diagonal approximation we have $\nu = (n,q)$ and
\begin{equation}
|\,z^{\nu} \rangle = \zeta^{\vphu}_{n,q}\,|\,\ZRPA^{n} \rangle\,,\quad
\omega^{\vphu}_{\nu} = \tilde{\omega}^{\vphu}_{n,q}\,,\quad
\sigma^{\vphu}_{\nu} = \sigma^{\vphu}_{n}\,.
\label{cbarentba}
\end{equation}
From Eqs. (\ref{zzrpa}) and (\ref{cbarentba})
it follows that in this case we have
\begin{equation}
(z^{\nu})^2_{\mbsu{RPA}} = \zeta^2_{n,q}\,.
\label{zzrpaz2}
\end{equation}
Altogether, condition (\ref{czz2}) in the diagonal approximation
takes the form
\begin{equation}
\zeta^2_{n,q} > \zeta^2_{\mbsu{min}} = 1/2\,.
\label{czetda}
\end{equation}

After all, we can finally argue in favor of the
choice $\zeta^2_{\mbsu{min}}$ for the blocking criterion:
From Eq. (\ref{ansum}) we see that
\\
(i) there exists not more than one pole of the function
${\tilde R}^{\mbss{CBA}}_{nn}(\omega)$ for which
$\zeta^2_{n,q} > 1/2$, and
\\
(ii) two and more poles of this function can satisfy the condition
$\zeta^2_{n,q} > \zeta^2_{\mbsu{min}}$ in case of
$\zeta^2_{\mbsu{min}} < 1/2$.
\\
Thus one can consider the value $\zeta^2_{\mbsu{min}} = 1/2$
as a threshold below which the fragmentation of the RPA state
$|\,\ZRPA^{n} \rangle$ becomes significant.
If there are no poles of the function ${\tilde R}^{\mbss{CBA}}_{nn}(\omega)$
with $\zeta^2_{n,q} > 1/2$,
then all the TBA-fragments of the RPA state $|\,\ZRPA^{n} \rangle$
have the structure going beyond the $1p1h$ approximation.
This conclusion corroborates the reasoning used in the derivation
of the condition (\ref{czz2}).

In the calculations presented below we use the CBA scheme in the
diagonal approximation. In this scheme, the {\it active} TBA phonons
(see Sec.~\ref{sec:CBA}) are found from
the solution of the system of the equations (\ref{poleq3}) together with
(\ref{andef}) and (\ref{cbarentba}).
According to the last equation, this scheme can be also called renormalized TBA.
Eq. (\ref{tbaeve}) is solved
in CBA using the full $1p1h$ space, i.e. without diagonal
approximation.  To solve the system (\ref{poleq3}), an iterative
procedure is employed with the using an exclusion method in which
the space of the active TBA phonons may only decrease starting from
a certain iteration.  This provides eventually a convergent procedure.

\subsection{Construction of the space of the RPA phonons}
\label{sec:RPAspace}

It is the strength of CBA that it implies a natural criterion for the
selection of those phonons which are active in the induced
interaction $W$. There remains, nonetheless, an issue of efficiency.
The diagonal approximation outlined in sections
\ref{sec:CBAdiag} and \ref{sec:CBAjust}
proceeds through a representation in terms of RPA phonons and these
have much different impact on $W$. Thus it is useful to restrict the
summation to the most important phonons. This is, at a technical
level, again the same quest for finding the most collective phonons as
in standard TBA and many different recipes are used in the literature
using a phonon coupling model.
In connection with standard TBA, we had introduced
in Ref. \cite{optim2017} an efficient criterion
for the selection of the collective RPA phonons.
The idea is to take the average strength
$\langle\,V\,\rangle^{\vphu}_n$
of the RPA residual interaction in state $n$ as measure of
collectivity. This is plausible because it is the residual
interaction which mixes the pure $1p1h$ states to a coherent
superposition of many states. Moreover, states with large
$\langle\,V\,\rangle^{\vphu}_n$ are generally strong coupling states
and thus will also contribute dominantly to the induced interaction.
Considering this strength relative to excitation energy then
leads
to
 the dimensionless measure of collectivity
\begin{subequations}
\begin{eqnarray}
  v^{\vphu}_n
  &=&
  \langle\,V\,\rangle^{\vphu}_n /\,|\,\omega^{\vphu}_{n}|\,,
\label{vndef}
\\
  \langle\,V\,\rangle^{\vphu}_n
  &=&
  \langle\,\ZRPA^{n}\,|\,V\,|\,\ZRPA^{n} \rangle
\nonumber
\\
  &=&\!\!
  \sum_{ph}
  [(\omega^{\vphu}_{n}\!-\!\ve^{\vphu}_{ph})|\ZRPA^{n}_{ph}|^2
  -(\omega^{\vphu}_{n}\!+\!\ve^{\vphu}_{ph})|\ZRPA^{n}_{hp}|^2]
  ,
\nonumber
\end{eqnarray}
where $\ve^{\vphu}_{ph}=\ve^{\vphu}_{p}-\ve^{\vphu}_{h}$.  We include
into the phonon basis of the TBA only the phonons with
\begin{equation}
  |\,v^{\vphu}_n\,| > v_{\mbss{min}}
\end{equation}
\end{subequations}
for some given value of $v_{\mbss{min}}$.  This criterion had been
tested extensively in \cite{optim2017}.  Plotting the distribution of
$v^{\vphu}_n$ we found a clear threshold value of $v^{\vphu}_n=0.05$
below which distribution becomes rapidly diffuse and we took this as a
physically sound cutoff criterion.

However, the criterion has two mild drawbacks. First, taking simply
the average residual interaction $\langle\,V\,\rangle^{\vphu}_n$
overlooks a few collective states having strong coupling matrix
elements which unfortunately happen to compensate each other in the
average. Second, the simple inverse energy weight gives much emphasis
to high energy phonons while we expect the strongest contributions to
$W$ from the low-energy collective modes. This leads us to propose
here an improved criterion based on the average of the square of
the residual interaction in state $n$, which can be reduced to
\begin{subequations}
\begin{eqnarray}
  \langle\,V^2\rangle^{\vphu}_n
  &=&
  \langle\,\ZRPA^{n}\,|\,V^2|\,\ZRPA^{n} \rangle
\nonumber
\\
  &=& \sum_{ph}\bigl[(\,\omega^{\vphu}_{n} - \ve^{\vphu}_{ph})^2
  |\ZRPA^{n}_{ph}|^2
\nonumber\\
  &&\qquad
  + (\,\omega^{\vphu}_{n} + \ve^{\vphu}_{ph})^2
  |\ZRPA^{n}_{hp}|^2 \bigr]
  \,,
\label{v2mn}
\end{eqnarray}
%
where the Hermitean property $V^{\dag}=V$ is taken into account.
From this, we form the dimensionless quantity
\begin{equation}
  \kappa_n = \frac{\gamma^2_n}{1+\gamma^2_n}
\label{kappdef}
  \;,\quad
  \gamma^2_n = \langle\,V^2\rangle^{\vphu}_n/\,\omega^2_n
\end{equation}
\end{subequations}
as new measure for collectivity.
Now we see that $\kappa_n=0$
requires that RPA eigenfrequency equals exactly
one of the $\pm\ve^{\vphu}_{ph}$ and thus is strictly uncorrelated.
Consequently, small values of $\kappa_n$ signal small
collectivity, i.e. coupling strength, throughout. Moreover, the
energy weight $\omega^{-2}_n$ yields the wanted weight on low-energy
states.

By noting that the amplitudes of the quasiparticle-phonon interaction
$g^{n}_{12}$ in the RPA are defined as
%
\be
g^{n}_{12} = \sum_{34} {V}^{\vphu}_{12,34}\,Z^{n}_{34}\,,
\label{grpadef}
\ee
we obtain from equations (\ref{v2mn}) and (\ref{grpadef})
\be
\langle\,V^2\rangle^{\vphu}_n = \langle\,{g}^{n}\,|\,{g}^{n} \rangle\,.
\label{v2mngg}
\end{equation}
In the macroscopic approach (see, e.g., Refs.~\cite{BM2,BBB83}) the
amplitudes $g^{n}_{12}$ are proportional to the deformation parameters
$\beta^{\vphu}_n$ of the respective vibrational modes.  So, in this
approach $\gamma^2_n \propto \beta^2_n$.  This shows explicitly that
the selection of the phonons with the largest values of $\gamma^2_n$
corresponds to the selection of the low-energy vibrational modes with
the largest deformation parameters having the most strong coupling to
the single-particle states.

Finally note that the states which are usually referred to as the
collective vibrational modes ($3^-_1$ levels in $^{16}$O and $^{40}$Ca,
$2^+_1$, $3^-_1$ and $4^+_1$ levels in $^{48}$Ca, $2^+_1$, $3^-_1$, $4^+_1$,
$5^-_1$ and $6^+_1$ levels in $^{208}$Pb) have $\gamma^2_n\gtrsim 1$
and, consequently, $\kappa_n \gtrsim 0.5$.  These archetype collective
modes  set the benchmark for collectivity. States with
$\kappa_n\approx 0.1$ are still acceptably strong phonon.  Values $\ll
1$ signal non-collective states.

\subsection{Standard TBA as limit of CBA}
\label{sec:TBAlim}

As stated above, standard TBA can be obtained as the first iteration
of CBA. It amounts to replacing the TBA spectrum in the equations
(\ref{eq:Winduc}) for the induced interaction $W$ by the mere RPA
spectrum, i.e. by simply identifying in the $1p1h\otimes$phonon
summations
\begin{equation}
 c^{CBA} = \{p',h',\nu\} \longrightarrow c^{RPA} = \{p',h',n\}
\end{equation}
where we emphasize the transition here symbolically through the
upper indices $CBA$ and $RPA$.

The discussion of cutoff criteria in the selection of RPA states, see
previous section, is of particular importance for standard TBA. A
proper selection of a few most collective states is compulsory in any
phonon coupling model to avoid double counting of complex
configuration and violation of the Pauli principle.

\section{Results and discussion}
\label{sec:res}

Before presenting the results, a few words about the calculations are
in order.  We have computed the same test cases with standard TBA and
with CBA in comparison using the same numerical procedures which are
explained in detail in
Refs.~\cite{optim2017,Tselyaev2016,Lyutorovich2017}.  The maximum
energy of the single-particle states of the $1p1h$ basis was taken as
100~MeV.  For $^{16}$O and $^{40}$Ca, we did not use an energy cutoff
on the phonon space while for $^{208}$Pb, the phonon basis was
restricted by a maximum phonon energy of 100~MeV. For all cases we
use the Skyrme parametrization SV-m64k6 \cite{Lyutorovich2012} which,
due its low effective mass, provides a particularly critical test of
the impact of stability and convergence.

It is standard practice to check the properties of standard TBA under
variation of the cutoff parameter, $v_n$ or $\kappa_n$
respectively. This is not so obvious in CBA because the proper cutoff
is set by theoretical considerations.  Nonetheless, we use the same
sort of cutoff in RPA space as a preselection of the expansion
basis. Thus one can very well present results as a function of this
preselection cutoff $\kappa_\mathrm{min}$ and so study convergence of
the method. This is what we will do in several of the following
figures.

In Fig. \ref{fige3f} we present the results of the energies of the first
$3^-$ levels in $^{16}$O, $^{40}$Ca, and $^{208}$Pb.
The value $\kappa_\mathrm{min}=1$ corresponds to the RPA.
The CBA results show nice convergence with the
expansion space $1/\kappa_\mathrm{min}$ and the converged result
agrees perfectly with the results of our previous analysis
\cite{optim2017}. The standard TBA result, however, shows a almost
constant slope (in logarithmic $\kappa_\mathrm{min}$ scale).  Here we
need again a separate analysis of the distribution of $\kappa_n$ to
find the optimum value of $\kappa_n$. Doing that we find an optimum
cutoff $\kappa_\mathrm{min}$
in the interval 0.05--0.1
which yields excitation energies
again close to the previous results (fine dashed horizontal line) and
CBA.  The interesting message is that CBA comes to the correct result
without separate decision on cutoff parameters. Of course, one still
wants to check convergence with $\kappa_n$. But this is a technical
aspect.

\begin{figure}
\centerline{\includegraphics[width=1.05\linewidth,trim={20mm 45mm 10mm 10mm},clip]{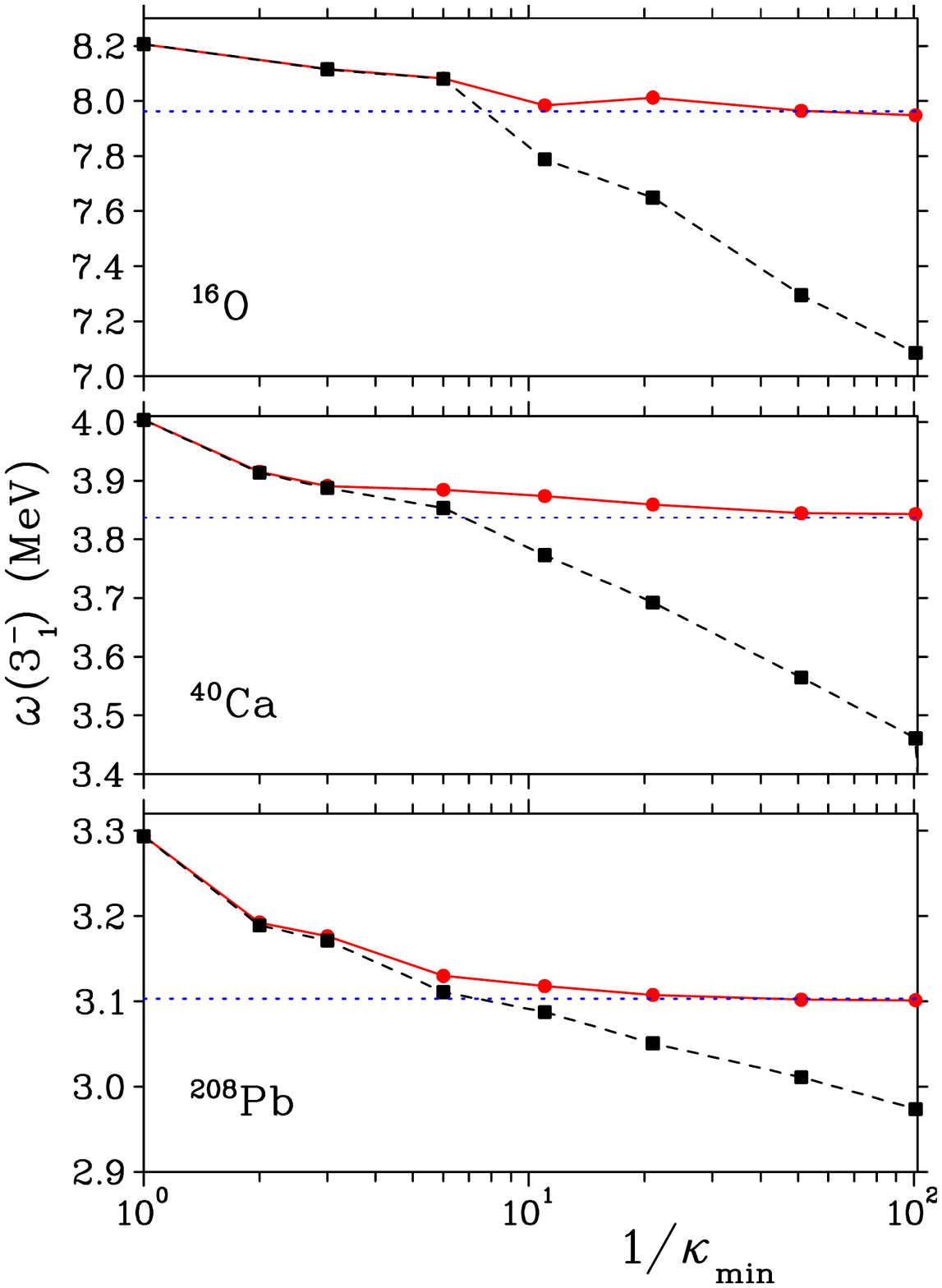}}
\caption{\label{fige3f} Dependence of the energy of $3^-_1$ state
  calculated in the CBA with $\zeta^2_{\mbsu{min}} = 0.5$ (red
  solid lines) and in the standard TBA (black dashed lines) on the
  value of the inverse cutoff parameter $1/\kappa_{\mbsu{min}}$.  The
  fine dashed horizontal line indicates the energy as found in a previous
  paper using standard TBA with optimized cutoff parameter $v_{\mbsu{min}} = 0.05$
  \cite{optim2017}. The experimental values are given in Table~\ref{tab:e3f}.}
\end{figure}

\begin{figure}
\centerline{\includegraphics[width=1.2\linewidth]{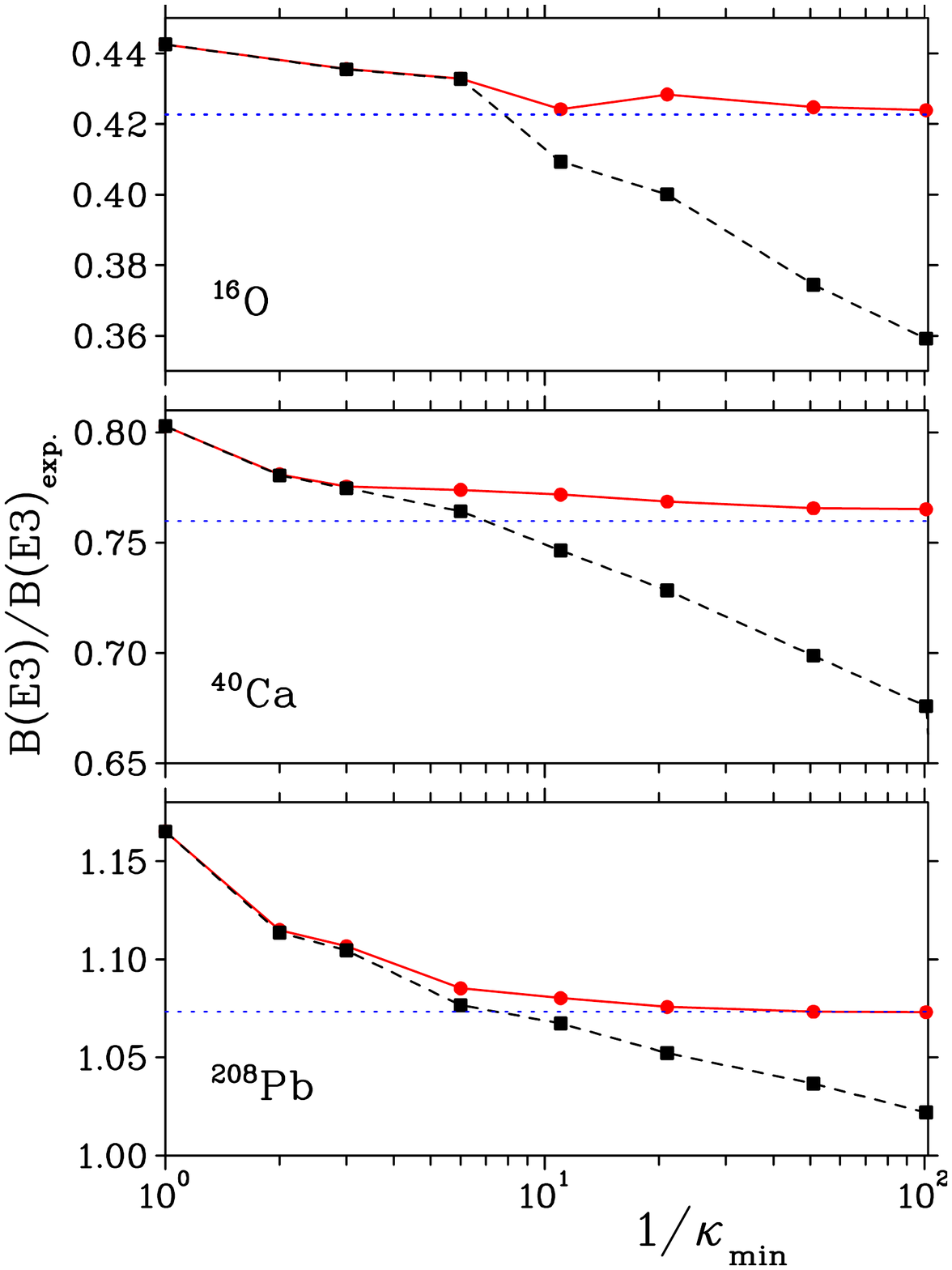}}
\vskip -3.00cm
\caption{\label{fige3f3BE3}
Same as in Fig. \ref{fige3f} but for the reduced probabilities
$B(E3; 0^+_{\mbsu{g.s.}} \rightarrow 3^-_1)$
in units of their experimental values.}
\end{figure}

 In Fig. \ref{fige3f3BE3} the transition probabilities of the $3^-_1$ states
 in these three nuclei are presented. The quantities  are the most sensitive
 properties for any nuclear structure model as they are directly connected
 with the collective wave function. First of all one notices that the values
 converge within our newly developed theory whereas the conventional TBA results
 do not show this behavior. For $^{208}$Pb the agreement with data is very good,
 for $^{40}$Ca fair but for $^{16}$O we are off by more then a factor of two.
 Obviously in light nuclei one
does not have enough $1p1h$ configurations for creating collective states.

\begin{figure}
\centerline{\includegraphics[width=1.05\linewidth,trim={20mm 65mm 10mm 10mm},clip]{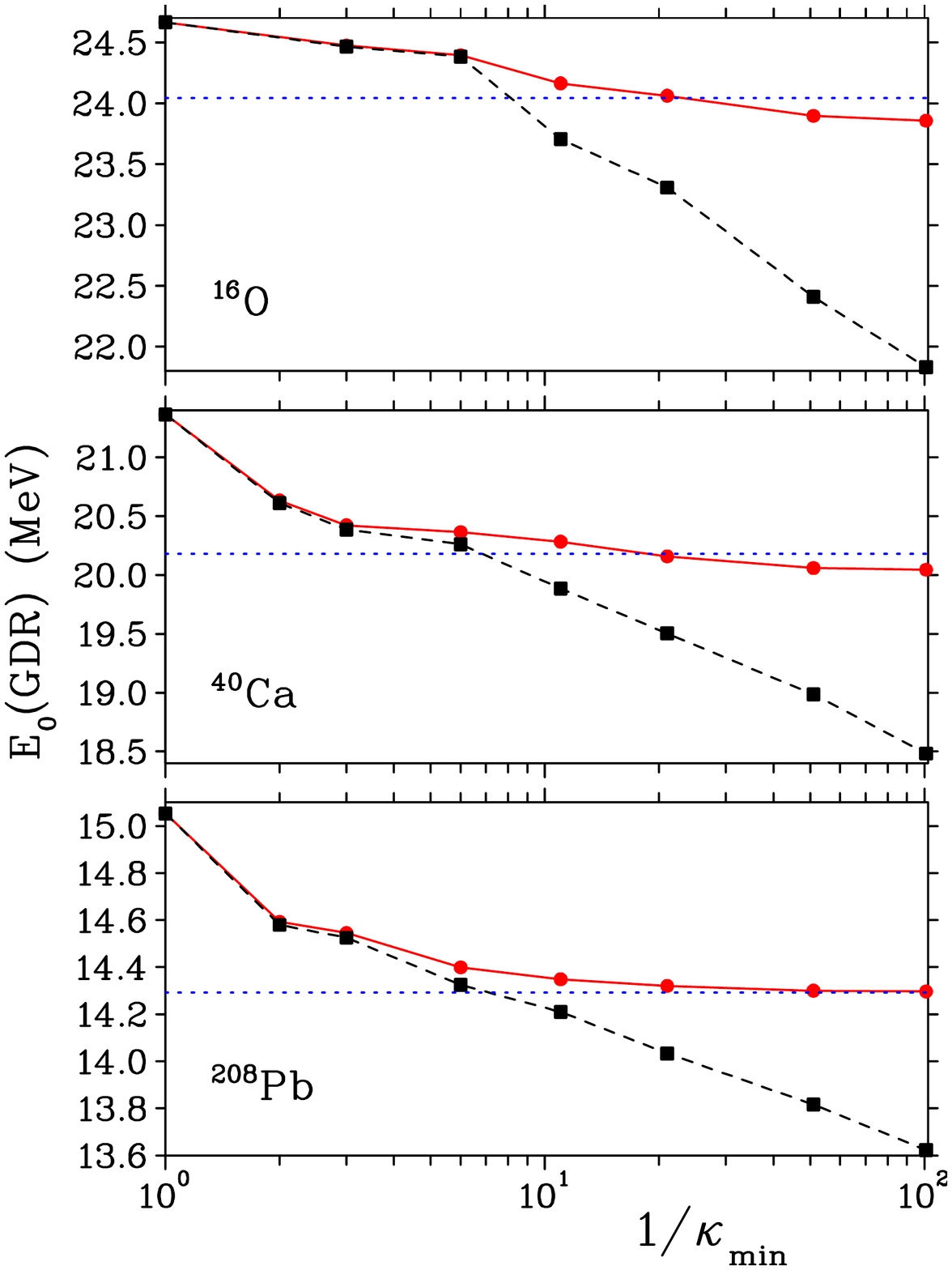}}
\caption{\label{fige1f} Same as in Fig. \ref{fige3f} but for the mean
  energies (Lorentzian parameter $E_0$) of the giant dipole
  resonance.}
\end{figure}

Fig.~\ref{fige1f} shows the analogous results for the mean energies of
the giant dipole resonance (GDR). The mean energies were defined as
the values of the Lorentzian parameter $E_0$ determined by equating
the energy-weighted moments of the calculated strength functions with
the respective moments of the Lorentzian function. The moments were
calculated within the following finite energy intervals whose centers
approximately coincide with $E_0$: 0--40 MeV for $^{16}$O, 10--30 MeV
for $^{40}$Ca, and 7--21 MeV for $^{208}$Pb. The trends, observations,
and conclusions are exactly the same as in the previous figures.

\begin{figure}
\centerline{\includegraphics[angle=90,height=0.78\linewidth,trim={23mm 53mm 10mm 15mm},clip]{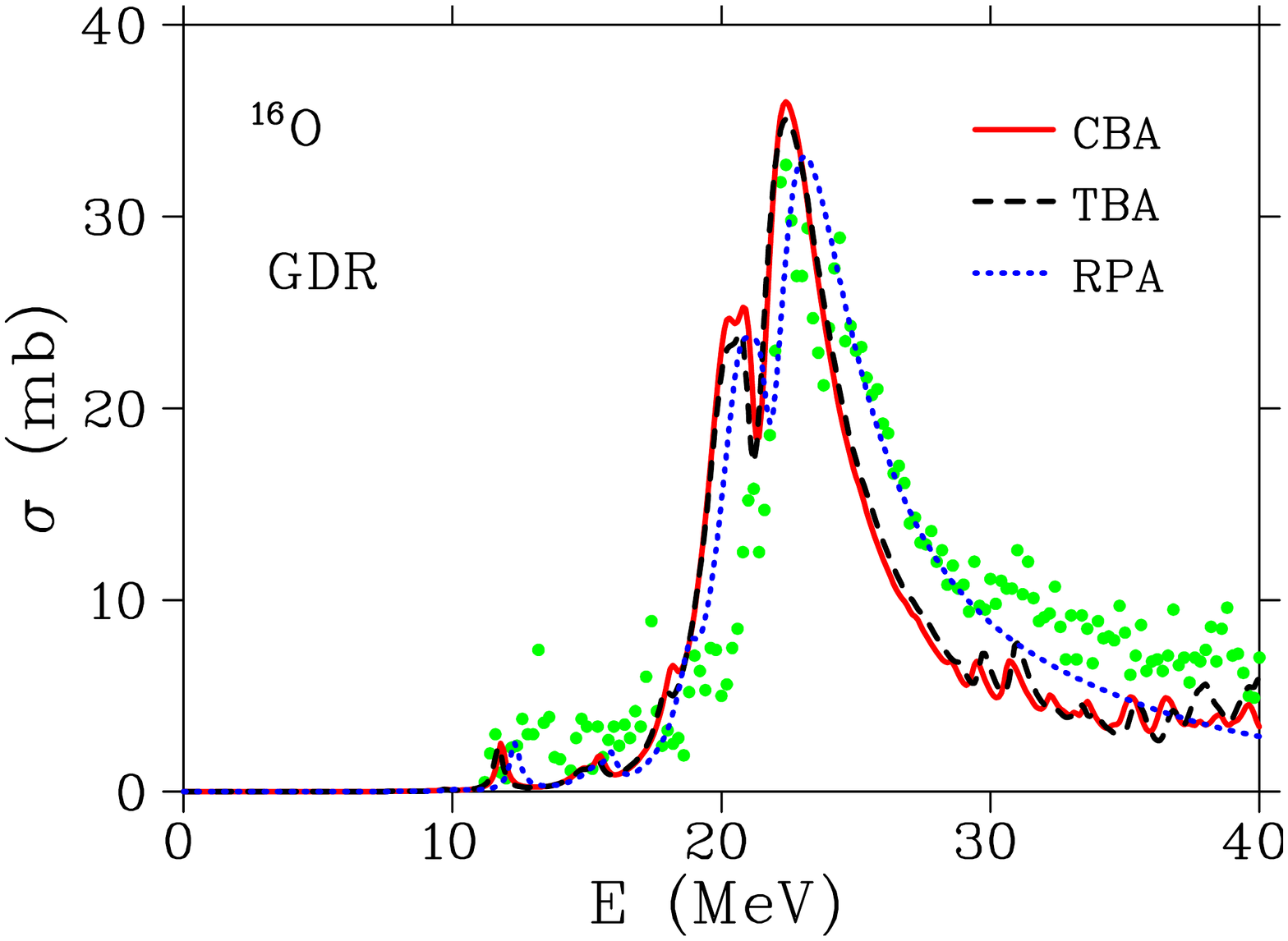}}
\caption{\label{o16gdr} The giant dipole resonance (GDR) in $^{16}$O
  calculated the CBA (red solid line), in standard TBA (black
  dashed line), and in RPA (blue dotted line). Experimental data from
  Ref.~\cite{o16GDRdata} are shown by the green circles.}
\end{figure}

\begin{figure}
\centerline{\includegraphics[angle=90,height=0.78\linewidth,trim={23mm 53mm 10mm 15mm},clip]{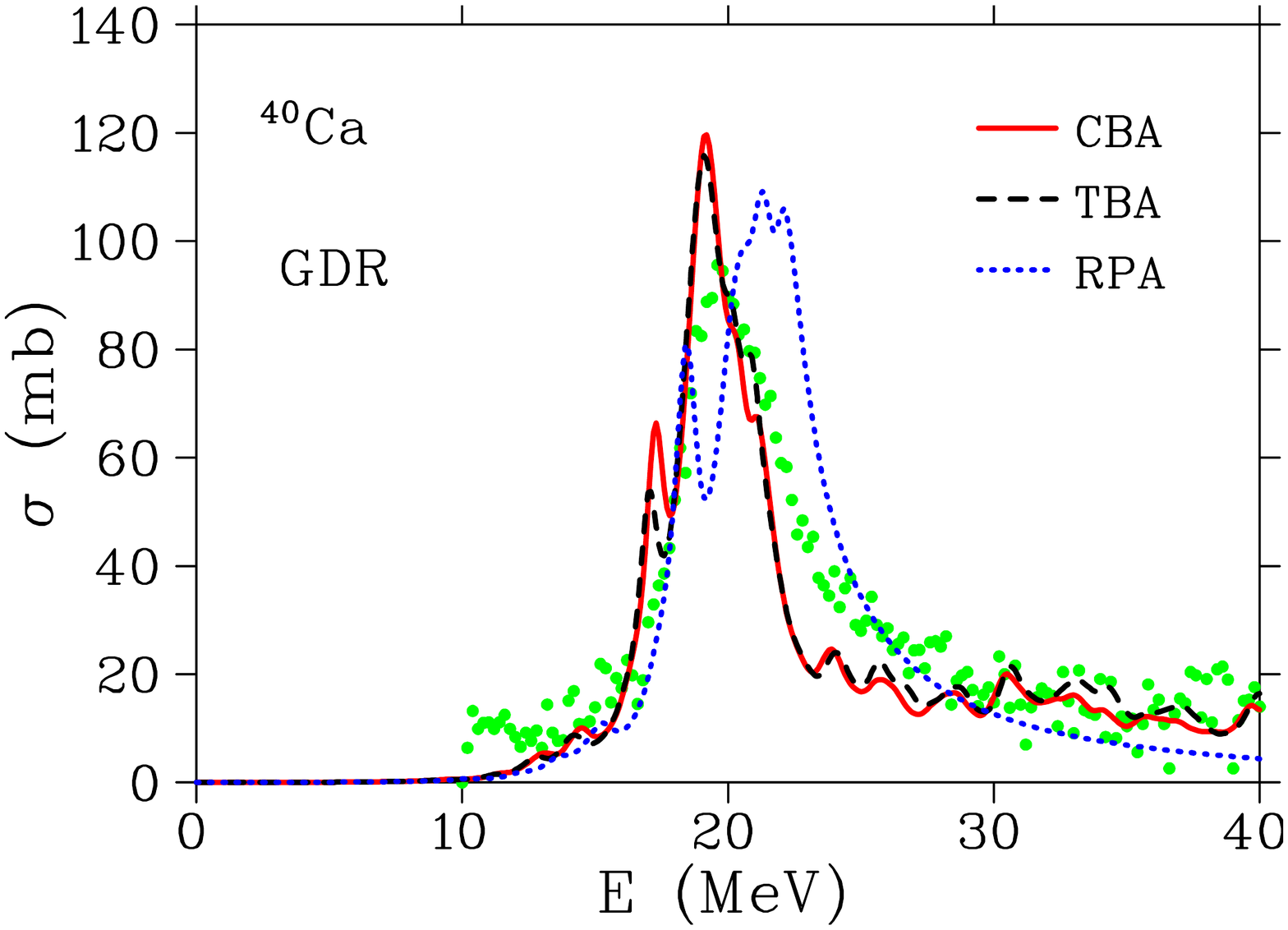}}
\caption{\label{ca40gdr}
Same as in Fig. \ref{o16gdr} but for $^{40}$Ca.
Experimental data are taken from Ref.~\cite{ca40GDRdata}.}
\end{figure}

\begin{figure}
\centerline{\includegraphics[angle=90,height=0.78\linewidth,trim={23mm 53mm 10mm 15mm},clip]{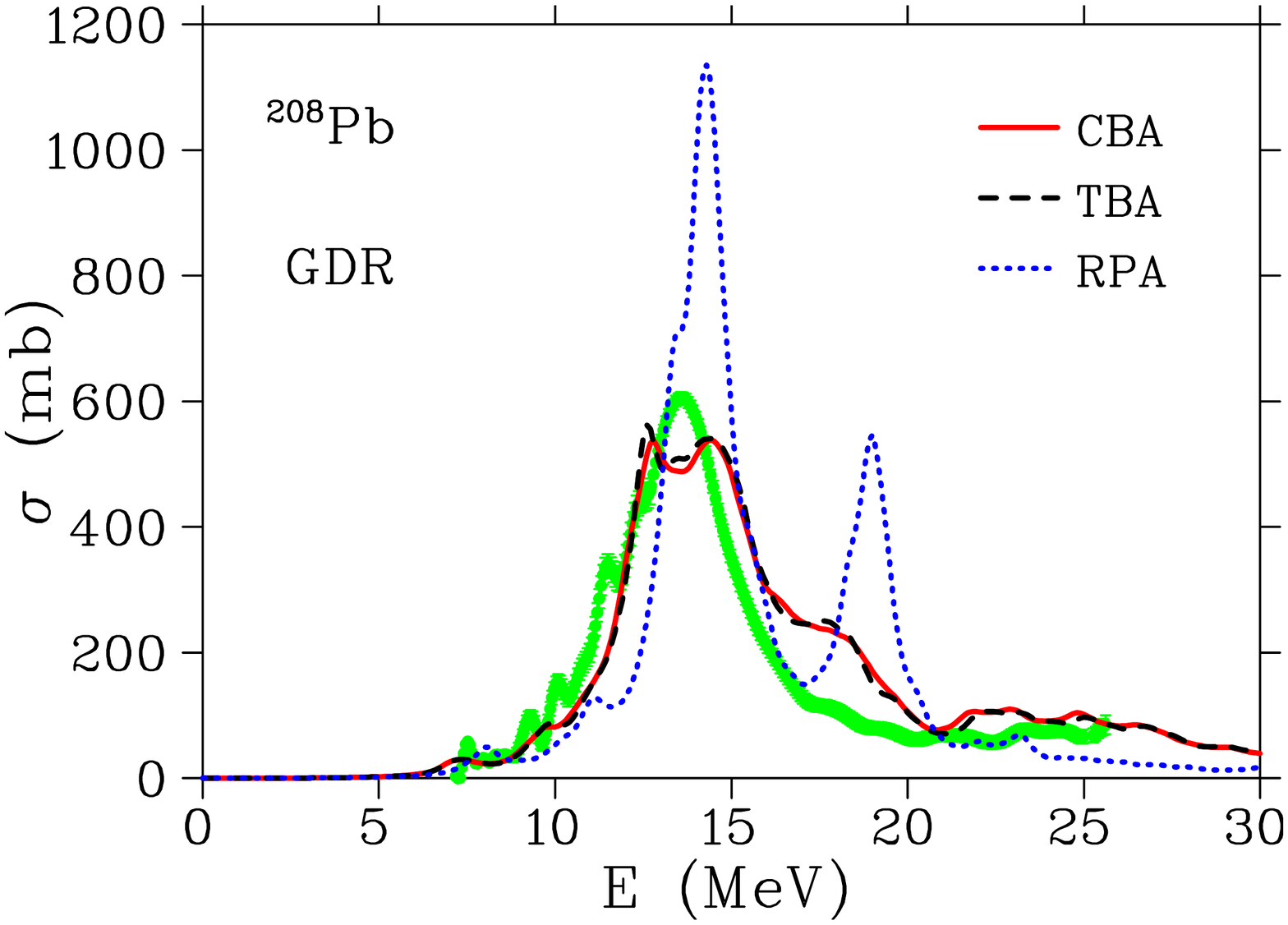}}
\caption{\label{pb208gdr}
Same as in Fig. \ref{o16gdr} but for $^{208}$Pb.
Experimental data are taken from Ref.~\cite{pb208GDRdata}.}
\end{figure}

The detailed distributions of the GDR strength distributions (the
photoabsorption cross sections) in $^{16}$O, $^{40}$Ca, and $^{208}$Pb
are shown in Figs. \ref{o16gdr}--\ref{pb208gdr}.  In all calculations
of the GDR strength in $^{16}$O and $^{40}$Ca the single-particle
continuum was included as described in \cite{Tselyaev2016}. The
smearing parameters were 200~keV in $^{16}$O and
400~keV in $^{40}$Ca and $^{208}$Pb. Standard TBA calculations which
were done for comparison use the optimized cutoff parameter $v_{\mbsu{min}}=0.05$.
 The value $\kappa_\mathrm{min}=0.01$ was
used in CBA for limiting the RPA
expansion basis which is well on the safe side as we see from Figs. \ref{fige3f}--\ref{fige1f}.

Figs. \ref{o16gdr}--\ref{pb208gdr} shows the same relative trends for
all three nuclei. The spectra from CBA agree well with those of
standard TBA and both differ significantly from RPA. TBA induces a
small down-shift and, more important, smooths the spectra
significantly.  This smoothing is important to bring the
theoretical distributions close to the experimental profile.

\begin{figure}
\centerline{\includegraphics[width=1.05\linewidth,trim={20mm 55mm 10mm 10mm},clip]{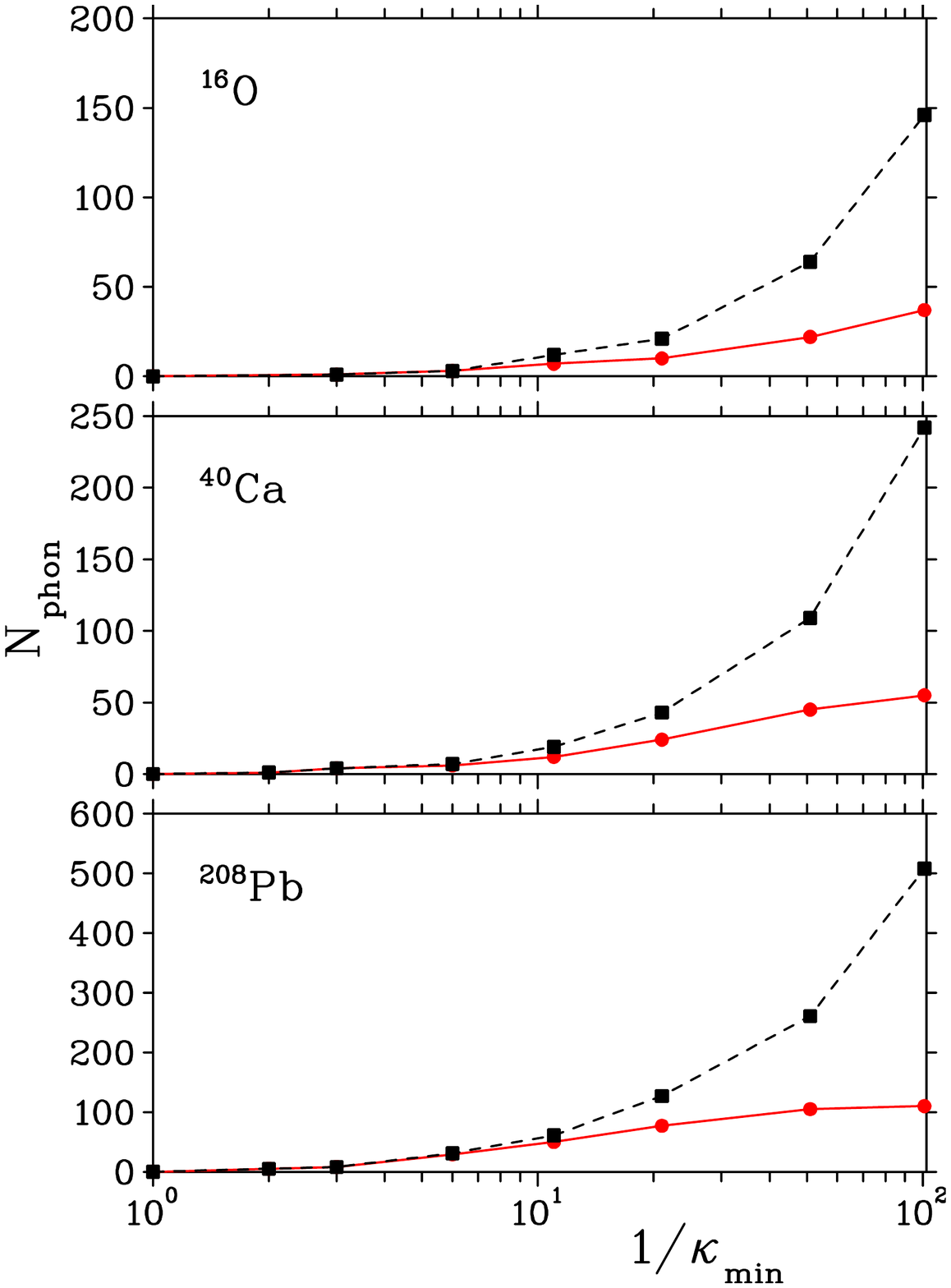}}
\caption{\label{fignphon}
Dependence of the number of the active TBA phonons in the CBA
with $\zeta^2_{\mbsu{min}} = 0.5$ (red solid lines)
and the number of the RPA phonons in the ordinary TBA
(black dashed lines) on the value of the cutoff parameter $\kappa_{\mbsu{min}}$.
}
\end{figure}

As can be seen from Figs. \ref{fige3f}--\ref{fige1f}, the
  convergence in the CBA results is achieved at $\kappa_{\mbsu{min}}
  \approx \gamma^2_{\mbsu{min}}$ in the interval $0.02-0.05$ which
  approximately corresponds to the boundary of the non-collective
  phonons as assessed by plotting the density of phonon states
    \cite{optim2017}.  The results of the ordinary TBA do not show
  the tendency to convergence.  The reason of the convergence in the
  CBA is a decrease of the number of phonons as compared with the
  ordinary TBA owing to an additional criterion for the selection of
  the phonons, Eq. (\ref{czetda}), determined by the parameter
  $\zeta^2_{\mbsu{min}}$. This decrease becomes very strong when
  $\gamma^2_{\mbsu{min}}$ and $\kappa_{\mbsu{min}}$ are very small
  (see Fig.~\ref{fignphon}).  At $0.02 \leqslant \gamma^2_{\mbsu{min}}
  \leqslant 0.05$ the number of the active TBA phonons in the CBA
  is in the intervals 10--22 for $^{16}$O, 24--45 for $^{40}$Ca, and
  77--105 for $^{208}$Pb.  In practice, all the phonons with $\omega_n
  > \omega_{\mbsu{max}}$ for some value of $\omega_{\mbsu{max}}$
  appear to be strongly fragmented, that is they have $\zeta^2_n <
  \zeta^2_{\mbsu{min}}=1/2$ and for this reason are excluded from the
  basis.  The value of $\omega_{\mbsu{max}}$ is different for
  different nuclei.  In the calculations presented in
  Figs. \ref{fige3f}--\ref{fignphon}, it is approximately 30 MeV for
  $^{16}$O, 26 MeV for $^{40}$Ca, and 14 MeV for $^{208}$Pb. It
  is highly satisfying to see that the new selection criterion
  provides the same cutoff in phonon space as the previously external
  cutoff criterion developed from the density of phonon states
  \cite{optim2017} while the new criterion is now inherent in the
  scheme and so more natural.

The additional effect of the renormalization,
Eq. (\ref{cbarentba}), consists in a decrease of the phonon's
energies, because, as a rule,
$\tilde{\omega}^{\vphu}_{n}<\omega^{\vphu}_{n}$ for positive
$\omega^{\vphu}_{n}$. In most cases,
this decrease improves the agreement with the experiment.  For the
nuclei considered in the present work, the calculated
and experimental
values of the energies of the $3^-_1$ states are listed in
Table~\ref{tab:e3f}.  As before, the Skyrme parametrization
  SV-m64k6 was used.  The CBA results were obtained with
  $\kappa_{\mbsu{min}} = 0.01$ and $\zeta^2_{\mbsu{min}} = 0.5$.
  The CBA results in the diagonal approximation CBA(D) imply the
  approximation described in sections \ref{sec:CBAdiag} and \ref{sec:CBAjust},
  that concerns the
  active TBA phonons used in the CBA (see Sec.~\ref{sec:CBA}).  As
  can be seen from this Table, the diagonal approximation keeps the
  deviation from the exact results for these states within 0.4~\%. The
  central question in our investigation concerns the convergence as a
  function of the size of the phonon space and the results are indeed
  very convincing.

\begin{table}
\caption{\label{tab:e3f}
The energies of the $3^-_1$ states
in $^{16}$O, $^{40}$Ca, and $^{208}$Pb calculated within the RPA,
the CBA, and the CBA in the diagonal approximation CBA(D).
The experimental data are given in the last column.}
\begin{ruledtabular}
\begin{tabular}{rcccc}
& \multicolumn{4}{c}{Energy of the $3^-_1$ state (MeV)} \\
& RPA & CBA & CBA(D) & experiment \\
\hline
$^{16}$O   & 8.21 & 7.95 & 7.94 & 6.13 \\
$^{40}$Ca  & 4.00 & 3.84 & 3.83 & 3.74 \\
$^{208}$Pb & 3.29 & 3.10 & 3.09 & 2.61 \\
\end{tabular}
\end{ruledtabular}
\end{table}

\begin{figure}
\centerline{\includegraphics[angle=90,width=1.28\linewidth]{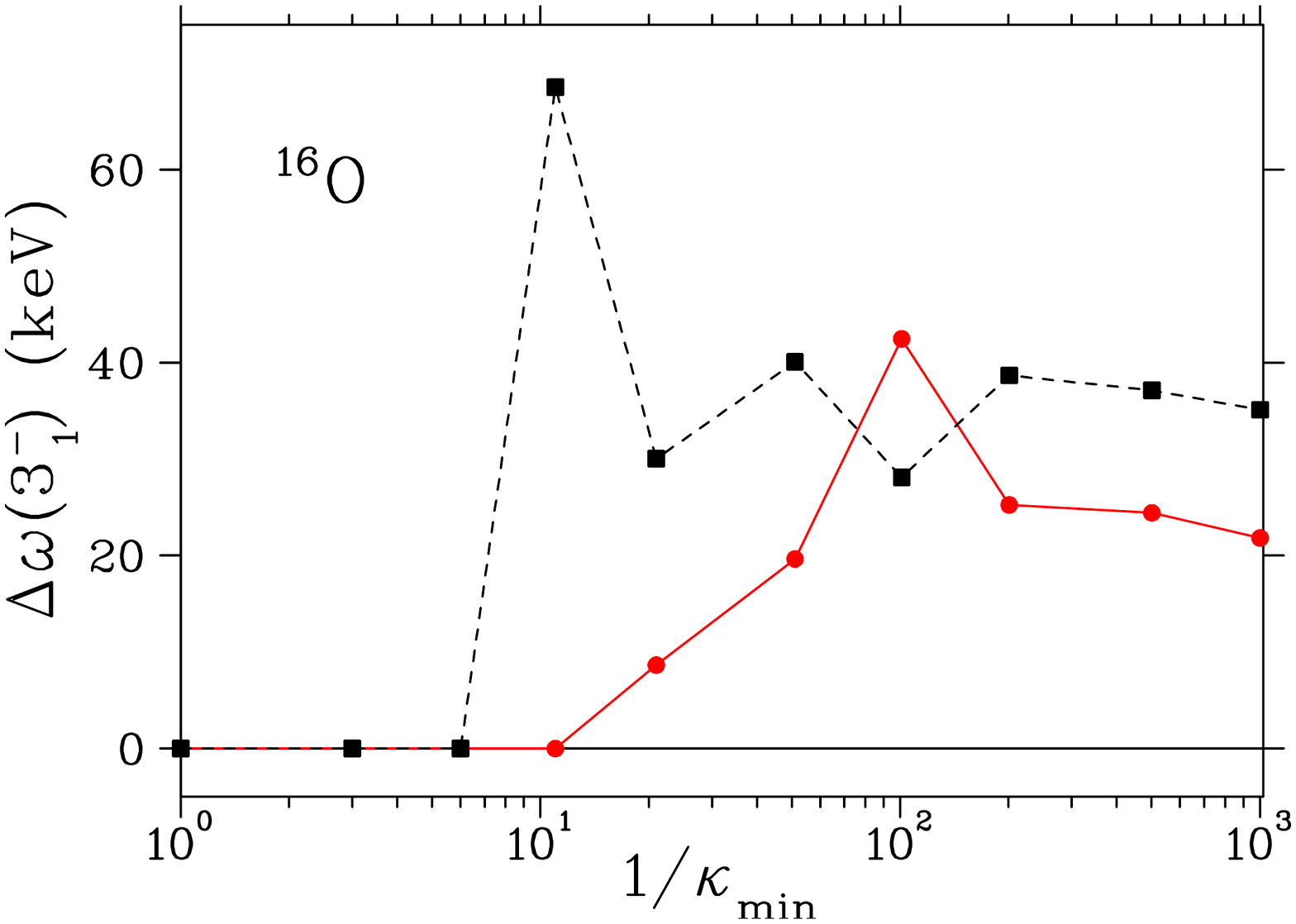}}
\caption{\label{fige3z2o16}
Differences between the energies of $3^-_1$ state
in the nucleus $^{16}$O as calculated in the CBA with
different values of the cutoff $\zeta^2_{\mbsu{min}}$,
namely
$\Delta \omega (3^-_1) =
\omega (3^-_1; \zeta^2_{\mbsu{min}} = 0.6) -
\omega (3^-_1; \zeta^2_{\mbsu{min}} = 0.5)$ (red solid lines) and
$\Delta \omega (3^-_1) =
\omega (3^-_1; \zeta^2_{\mbsu{min}} = 0.7) -
\omega (3^-_1; \zeta^2_{\mbsu{min}} = 0.5)$ (black dashed lines).
}
\end{figure}

\begin{figure}
\centerline{\includegraphics[angle=90,width=1.28\linewidth]{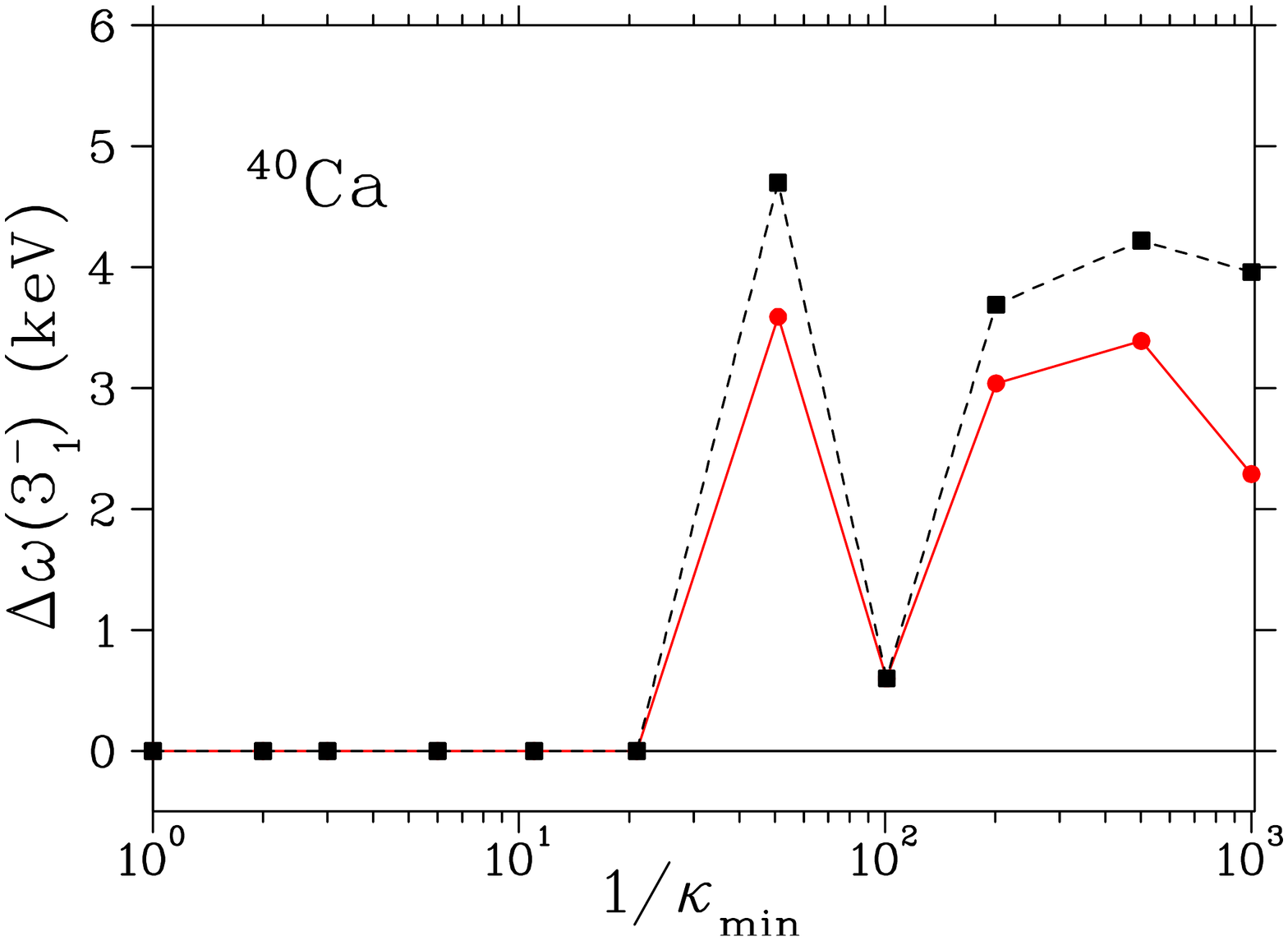}}
\caption{\label{fige3z2ca40}
Same as in Fig. \ref{fige3z2o16} but for the nucleus $^{40}$Ca.
}
\end{figure}

\begin{figure}
\centerline{\includegraphics[angle=90,width=1.28\linewidth]{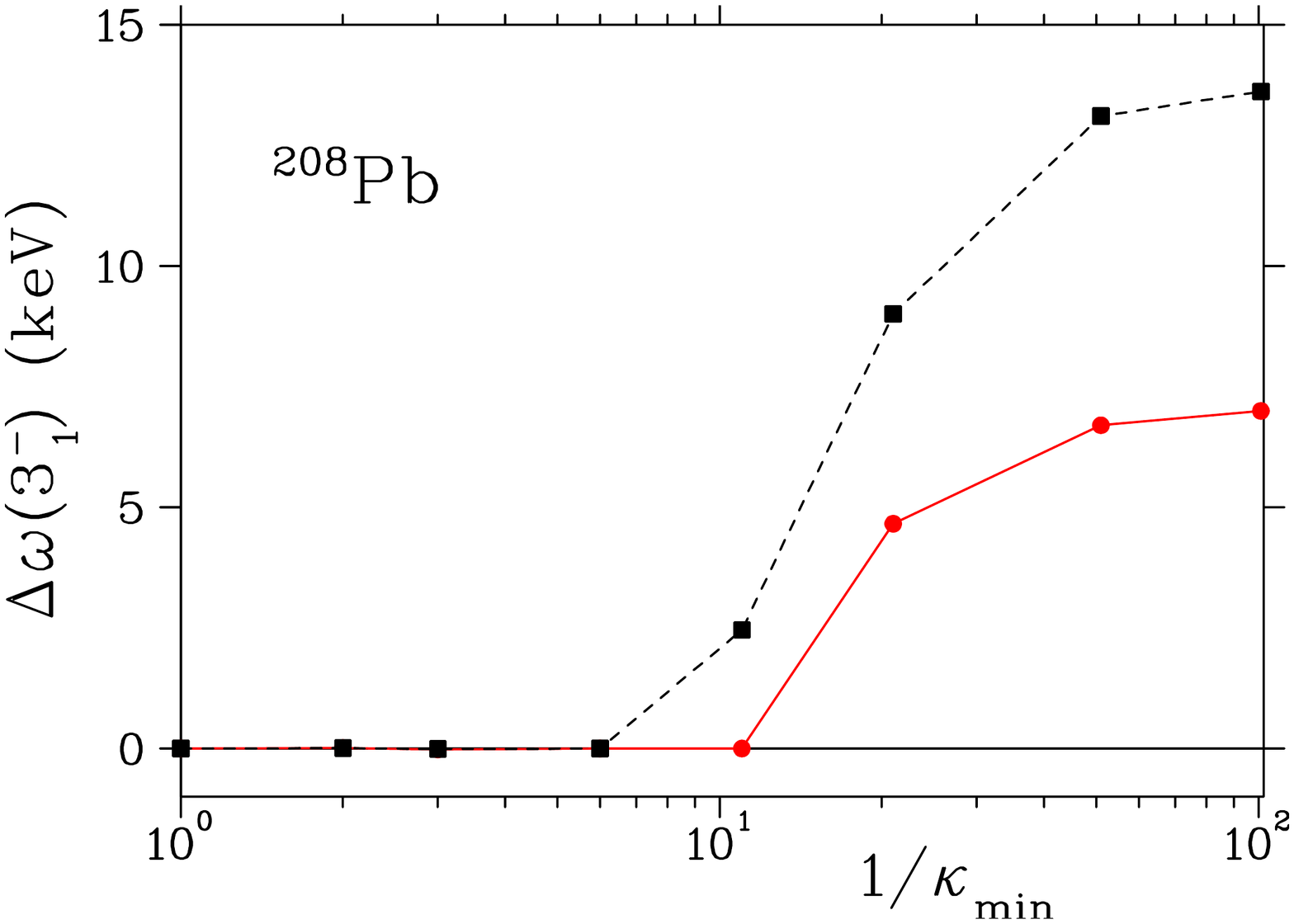}}
\caption{\label{fige3z2pb208}
Same as in Fig. \ref{fige3z2o16} but for the nucleus $^{208}$Pb.
}
\end{figure}

As mentioned above, the value $\zeta^2_{\mbsu{min}} = 0.5$
is a threshold below which the fragmentation of the RPA state becomes
significant. To study the sensitivity of the results to the increase of
$\zeta^2_{\mbsu{min}}$ we have calculated the energies of the first
$3^-$ levels in $^{16}$O, $^{40}$Ca, and $^{208}$Pb in the CBA with
$\zeta^2_{\mbsu{min}}$ equal to 0.6 and 0.7. The results are shown in
Figs. \ref{fige3z2o16}--\ref{fige3z2pb208} where the energy differences
$\Delta \omega (3^-_1) =
\omega (3^-_1; \zeta^2_{\mbsu{min}} = 0.6) -
\omega (3^-_1; \zeta^2_{\mbsu{min}} = 0.5)$ (red solid lines) and
$\Delta \omega (3^-_1) =
\omega (3^-_1; \zeta^2_{\mbsu{min}} = 0.7) -
\omega (3^-_1; \zeta^2_{\mbsu{min}} = 0.5)$ (black dashed lines)
are plotted versus the values of $1/\kappa_{\mbsu{min}}$.
As can be seen from Figs. \ref{fige3z2o16} and \ref{fige3z2ca40},
these energy differences have a tendency to decrease in $^{16}$O and $^{40}$Ca
in the region near $\kappa_{\mbsu{min}} = 10^{-3}$.
For $^{208}$Pb, they are stabilized in the region slightly above
$\kappa_{\mbsu{min}} = 10^{-2}$. The value of $\Delta \omega (3^-_1)$ at
$\zeta^2_{\mbsu{min}} = 0.6$ and $\kappa_{\mbsu{min}} = 10^{-2}$
amounts to 7~keV for $^{208}$Pb
(that corresponds to 4~\% of the overall shift of the CBA $3^-_1$ energy
with respect to the RPA one).
The values of $\Delta \omega (3^-_1)$ at
$\zeta^2_{\mbsu{min}} = 0.6$ and $\kappa_{\mbsu{min}} = 10^{-3}$
amount to 2~keV for $^{40}$Ca (1~\% of the overall shift) and
22~keV for $^{16}$O (8~\% of the overall shift).
Thus, one can conclude that the small increase of $\zeta^2_{\mbsu{min}}$
with respect to the value $\zeta^2_{\mbsu{min}} = 0.5$ only slightly
affects the results.

\section{Conclusions}
\label{sec:Conc}

We have studied phonon-coupling models for nuclear resonances within
the time blocking approximation (TBA). A generalized version of TBA is
developed in which the self-consistency principle is extended to the
phonon space of the model. This leads to a non-linear equation for the
energies and transition amplitudes of the nuclear excited states
(phonons).  The most general version of this non-linear equation is
simplified in two steps: First, the space of phonons to be included in
the expansion is limited by the natural requirement that only phonons
with dominant $1p1h$ contributions are
selected. This is the configuration blocking approximation (CBA) which
we use as name for the new scheme.  The formalism implies a precise
limit that the $1p1h$ content must be larger than 50\%. Second, one
invokes a diagonal approximation in the representation of the complete
set of the solutions of the RPA equations.  It turns out, that in this
diagonal approximation CBA is equivalent to the renormalization of the
amplitudes of the phonons entering the phonon basis of the model which
could also describe the new scheme alternatively as a renormalized
TBA.  The CBA is analyzed in the calculations of the first $3^-$
states and the giant dipole resonances in magic nuclei $^{16}$O,
$^{40}$Ca, and $^{208}$Pb.  It is shown that CBA produces
a natural convergence of the results with respect to enlarging the
phonon space of the model. This is an advantage as compared to
ordinary TBA where additional, external criteria are needed to limit
the phonon expansion basis in reasonable manner. This was done
previously by reading off the cutoff from the density of phonon
states. It is highly satisfying that the new, implicit cutoff in
CBA produces converged results in agreement with the previous
selection scheme.

\begin{acknowledgements}
V.T. and N.L. acknowledge financial support from the Russian Science
Foundation (project No.  16-12-10155).
This research was supported by the Computer Center of St. Petersburg State University.
This work has been supported by contract Re322-13/1 from the DFG.
\end{acknowledgements}

%

%
\end{document}